\documentclass[12pt,preprint]{aastex}

\usepackage{emulateapj5}
\submitted{Accepted for publication in the Astrophysical Journal}

\newcommand{\eg}{e.g., }
\newcommand{\ie}{i.e., }
\newcommand{\Msun}{M_{\odot}}
\newcommand{\kms}{km~s$^{-1}$}

\newcommand{\Fefs}{$^{56}$Fe}
\newcommand{\Cofs}{$^{56}$Co}
\newcommand{\Nifs}{$^{56}$Ni}

\newcommand{\Mej}{M_{\rm ej}}
\newcommand{\KE}{E_{\rm K}}
\newcommand{\Mni}{M(^{56}{\rm Ni})}
\def\gsim{\mathrel{\rlap{\lower 4pt \hbox{\hskip 1pt $\sim$}}\raise 1pt
\hbox {$>$}}}
\def\lsim{\mathrel{\rlap{\lower 4pt \hbox{\hskip 1pt $\sim$}}\raise 1pt
\hbox {$<$}}}

\def\ion#1#2{{\rm #1}~{\sc #2}}

\shorttitle{SN 2006gy at late phase}
\shortauthors{Kawabata et al.}
 
\begin{document}

\title{
The Extremely Luminous Supernova 2006\lowercase{gy} at Late Phase:
Detection of Optical Emission from Supernova
\altaffilmark{1}
}

\author{
Koji S. Kawabata\altaffilmark{2},
Masaomi Tanaka \altaffilmark{3,4},
Keiichi Maeda \altaffilmark{4,5},
Takashi Hattori \altaffilmark{6},
Ken'ichi Nomoto \altaffilmark{4,3},
Nozomu Tominaga \altaffilmark{7,3}, and
Masayuki Yamanaka \altaffilmark{2,8}
}

\altaffiltext{1}{Based on data collected at Subaru telescope, 
which is operated by the National Astronomical Observatory of Japan (NAOJ).}
\altaffiltext{2}{Hiroshima Astrophysical Science Center, Hiroshima University, Higashi-Hiroshima, Hiroshima, Japan; kawabtkj@hiroshima-u.ac.jp}
\altaffiltext{3}{Department of Astronomy, Graduate School of Science, 
University of Tokyo, Bunkyo-ku, Tokyo, Japan; 
mtanaka@astron.s.u-tokyo.ac.jp, nomoto@astron.s.u-tokyo.ac.jp,
tominaga@astron.s.u-tokyo.ac.jp}
\altaffiltext{4}{Institute for the Physics and Mathematics of the
Universe, University of Tokyo, Kashiwa, Japan;
maeda@ea.c.u-tokyo.ac.jp}
\altaffiltext{5}{Max-Planck-Institut f\"{u}r Astrophysik, 
Garching bei M\"{u}nchen, Germany}
\altaffiltext{6}{Subaru Telescope, NAOJ, Hilo, HI, USA; hattori@subaru.naoj.org}
\altaffiltext{7}{Optical and Infrared Astronomy Division, NAOJ, Osawa, Mitaka, Tokyo, Japan}
\altaffiltext{8}{Department of Physical Science, School of Science,
Hiroshima University, Higashi-Hiroshima, Hiroshima, Japan; 
myamanaka@hiroshima-u.ac.jp}

\begin{abstract}

Supernova (SN) 2006gy is an extremely luminous Type IIn SN characterized 
by the bright peak magnitude $M_{R}\sim$ $-$22 mag and its long duration.
The mechanism giving rise to its huge luminosity is still unclear.
We performed optical spectroscopy and photometry of SN 2006gy 
at late time, $\sim$ 400 days after the explosion, with the 
Subaru/FOCAS in a good seeing condition.
We carefully extracted the SN component, although there is an ambiguity
because of the contamination by bright nucleus of the host galaxy.
We found that the SN faded by $\sim$ 3 mag from $\sim$ 200 to 
$\sim$ 400 days after the explosion (i.e., by $\sim$ 5 mag from
peak to $\sim$ 400 days) in $R$ band.
The overall light curve is marginally consistent with the 
\Nifs\ heating model, although the flattening around 200 days
suggests the optical flux declined more steeply between $\sim$ 200 
and $\sim$ 400 days.
The late time spectrum was quite peculiar among all types of SNe.
It showed many intermediate width ($\sim 2000$ \kms\ FWHM) 
emission lines, e.g., [\ion{Fe}{ii}], [\ion{Ca}{ii}], and \ion{Ca}{ii}.
The absence of the broad [\ion{O}{i}] 6300, 6364 line and weakness of 
[\ion{Fe}{ii}] and [\ion{Ca}{ii}] lines compared with \ion{Ca}{ii}
IR triplet would be explained by a moderately high electron density 
in the line emitting region.
This high density assumption seems to be consistent with the large 
amount of ejecta and low expansion velocity of SN 2006gy.
The H$\alpha$ line luminosity was as small as 
$\sim 1\times 10^{39}$ erg s$^{-1}$, being comparable with those
of normal Type II SNe at similar epochs.
Our observation indicates that the strong CSM interaction had 
almost finished by $\sim$ 400 days.
If the late time optical flux is purely powered by radioactive decay, 
at least $\Mni\sim 3\Msun$ should be produced at the SN explosion.
In the late phase spectrum, there were several unusual emission lines 
at 7400--8800 \AA\ and some of them might be due to Ti or Ni 
synthesized at the explosion.

\end{abstract}


\section{Introduction}
\label{sec:intro}

Supernova (SN) 2006gy is an extremely luminous SN found by
\citet{qui06} near the nucleus of the host galaxy NGC 1260 
[$\mu = (m-M) =34.45$ mag].
Its absolute magnitude reaches $\sim$ $-22$ mag
(Ofek et al. 2007; Smith et al. 2007),
being more luminous than typical Type Ia supernovae
by a factor of $\sim 10$.
The SN was referred to as ``the most luminous SN'' at that time
\footnote{When only the peak magnitude is concerned, 
it has been known that SNe 2005ap and 2008es are more luminous 
than SN 2006gy (Quimby et al. 2007; Miller et al. 2008; 
Gezari et al. 2008).}.
In addition to the high peak luminosity, the light curve 
(LC) of SN 2006gy evolves very slowly, peaking 
at $t \sim$ 70 days after the explosion
(hereafter, $t$ means time after the explosion)
\footnote{We assumed the explosion date of SN 2006gy to be
2006 Aug 20 (Smith et al. 2007).}.
The total radiation energy exceeds $1 \times 10^{51}$ ergs, 
i.e., comparable to the kinetic energy of typical SNe.
The optical spectra of SN 2006gy show Type IIn-like features, 
\ie narrow emissions of H Balmer lines.

For a model of SN 2006gy, a `Type IIa' scenario 
(hybrid of Type IIn and Ia SN, like SNe 2002ic and 1997cy; 
Deng et al. 2004), \ie Type Ia SN in a dense circumstellar 
medium (CSM), has been suggested.
However, the radiation energy of SN 2006gy is comparable to 
(or overwhelming) the kinetic energy of Type Ia SNe so that 
this scenario seems unlikely (Ofek et al. 2007; Smith et al. 2007).
The progenitor of SN 2006gy should be a massive star.

The source of the huge optical luminosity is under debate.
Possible mechanisms are (i) thermal 
emission from hydrogen recombination front, (ii) interaction between 
CSM and SN ejecta, and (iii) radioactive decay of \Nifs\ .
Ofek et al. (2007) suggested that the luminosity comes from 
the CSM interaction, as in Type IIn/IIa SNe.
On the contrary, Smith et al. (2007) suggested that radioactive decay
is the primal (or least problematic) heating source, because
only weak X-rays were detected from SN 2006gy (suggesting
CSM interaction was not strong) and the observed expansion velocity 
of hydrogen envelope was unusually slow (suggesting too small
emitting radius for thermal radiation).
Agnoletto et al. (2009) gave an idea that the massive CSM was 
clumpy and both the CSM interaction and the radioactive decay
contributed to the enormous luminosity.
Furthermore, Smith et al. (2007) suggested that 
SN 2006gy is a pair-instability SN (PISN) because a huge amount of 
\Nifs\ ($> 10 \Msun$) is required to explain the peak luminosity.
However, Nomoto et al. (2007) shows that the LC of the PISN model
evolves too slowly to reproduce the observed LC.
Umeda \& Nomoto (2008) explored a possibility that 
core-collapse SNe could synthesize such a large amount of \Nifs.

Woosley, Blinnikov \& Heger (2007) showed that the collision
between the shells ejected by pulsational pair instability 
in massive stars initially having $\sim 90 \Msun$
could reproduce the peak luminosity of SN 2006gy.
Smith \& McCray (2007) also presented a simple ``shell-shocked'' model 
involving shock-deposited energy as the heat source.
In these two models, the shell is so thick that X-rays are absorbed.
Portegies Zwart \& van den Heuvel (2007) suggested that 
the SN 2006gy was an outcome of merging between two massive stars
in a dense and young cluster, being consistent with presence
of a large amount of hydrogen.

Recently, Smith et al. (2008a) presented results of late phase 
observations ($t=362$--$468$ days). The SN was clearly seen in NIR images
obtained with the adaptive optics on the Keck II telescope, while
any optical observation failed to detect the SN component
even for H$\alpha$ emission.
The NIR flux would be attributed to newly-formed dust particles 
or an IR echo.
Agnoletto et al. (2009) also successfully detected this SN 
in $K$ band at $t=411$--$510$ days, but failed to detect the SN in the
optical bands at $t=389.5$--$423.5$ days.

In this paper, we report on the results of our late time observation 
($t=394$ days) of SN 2006gy with the Subaru telescope.
The late phase luminosity and spectra may give clues to distinguish 
the underlying scenarios as also mentioned by Smith et al. (2008a).
In \S \ref{sec:obs}, we describe our observations and data reduction.
In \S \ref{sec:LC}, the LC of SN 2006gy is compared with those of
other SNe and results of model calculations.
In \S \ref{sec:spec}, the spectra of SN 2006gy are analyzed. 
Features in the late phase spectrum is further investigated by comparing
it with spectra of SNe of various types (\S \ref{sec:discussion}).
Finally we give conclusions in \S \ref{sec:conclusions}.

\section{Observations and Data Reduction}
\label{sec:obs}

The observations were carried out with the Faint Object Camera
And Spectrograph (FOCAS, Kashikawa et al. 2002).
FOCAS is an optical versatile instrument attached to the
Cassegrain focus of the 8.2-m Subaru telescope.
The images were recorded on two MIT/LL CCDs (2k$\times$4k, 
15 $\mu$m pixel$^{-1}$).
To minimize the harmful effect due to atmospheric dispersion in 
observation at low altitude, we used the Atmospheric 
Dispersion Corrector (ADC) in front of FOCAS, which reduces
the chromatic elongation to less than 0\farcs 1 within wavelengths
3500--11000 \AA\ at airmass $\leq 2.0$.
We took all images at airmass less than 1.9 (Table \ref{tbl:obs}),
which promises that the effect was negligible in our observations.

For spectroscopy, we obtained high resolution spectra 
($R\simeq 3600$) on 2006 Dec 25.4 ($t=127$ days) and 2007 Jan 24.4 
($t=157$ days) and a low resolution spectrum ($R\simeq 660$) 
on 2007 Sep 18.5 ($t=394$ days).
For the high resolution spectroscopy, we used a $0\farcs 4$ 
width slit, a 665 lines mm$^{-1}$ volume-phase holographic grism 
and SY47 order-cut filter, which gives a wavelength coverage
of 5300--7700 \AA\ and a spectral resolution of $\simeq 1.8$\AA\ .
For the low resolution one, we used a $0\farcs 8$ width 
offset-slit, a 300 lines mm$^{-1}$ blue grism and SY47 
filter, giving a wavelength coverage of
4800--9000 \AA\ and a spectral resolution of $\simeq 9.6$\AA\ .
The total exposure time was 1200 s on each night.
The direction of the slit was set to position angle (PA)
$-75\arcdeg$ to take spectra of both the nucleus of the host galaxy
NGC 1260 and the SN simultaneously.
Flux calibration and the correction for atmospheric absorption 
bands were performed using spectra of a spectrophotometric standard 
star obtained on the same night; BD $+28\arcdeg 4211$ in 2006 Dec,
Feige 34 in 2007 Jan and G191B2B in 2007 Sep.
The log of observation is shown in Table {\ref{tbl:obs}}.

For photometry, we took $V$ and $R$ band images just before the 
spectroscopy.
The exposure times of $V$ and $R$ bands are 3 s and 4 s in 2006 Dec,
5 s and 5 s in 2007 Jan and 10 s and 10 s in 2007 Sep, respectively.
We used $2\times 2$ on-chip binning in the CCD readout and the 
resultant scale was 0\farcs 208 per a binned-pixel.
The magnitude of the SN is calibrated with Landolt field stars
(Landolt 1992).

\subsection{Background subtraction}
\label{sec:backsub}

Since the SN was very close to the bright nucleus of the host galaxy
($\Delta\simeq 0\farcs 9$), the subtraction of the background 
component should be carefully performed, especially for
late phase data.

In the earlier phase data,
the SN was sufficiently bright and the artificial error 
originated in the adopted method for the background 
subtraction is not significant.
For photometry, we subtracted PSF-matched 394 days image
in the same band as the background, and then performed 
PSF-photometry.
Although this method gives an over-subtraction due to a
non-negligible SN component at $t=$ 394 days, the error should be 
small ($\sim 0.03$ mag) because the SN faded by about 4 mag
between $t=$ 127 days and 394 days.
The photometric error was derived from the root sum square
of the magnitude transformation error using the Landolt field stars
and the photon noises of the SN and the background.
For spectroscopy, we assumed that the two-dimensional (2D)
spectral image of the host galaxy is symmetric 
across the nucleus, and simply subtracted the mirrored 
spectral image across the nucleus. 

\begin{figure}
\begin{center}
\includegraphics[scale=1.0]{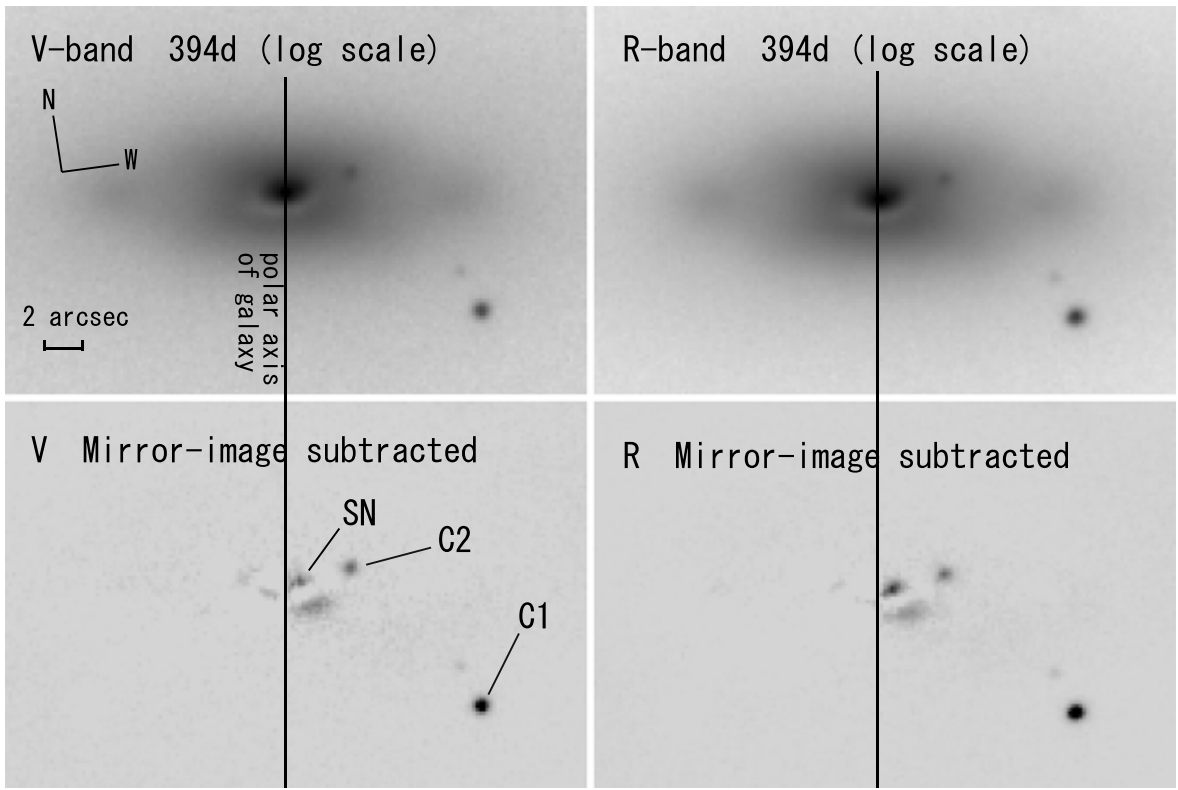}
\figcaption{
(Upper) $V$ and $R$ band images obtained by Subaru/FOCAS at $t=394$ days.
The images are rotated to align the polar axis of the host 
galaxy (PA$=-8\arcdeg$) to the vertical axis as shown.
(Lower) Background-subtracted images in $V$ and $R$ bands.
We assumed that the mirror image across the galactic polar axis
(i.e., the flipped left-side image) as the background.
We derived $V=20.7\pm 0.4$ and 
$R=19.4\pm 0.4$ for the SN (see \S\ref{sec:backsub}), $V=19.48\pm 0.14$ and 
$R=18.37\pm 0.13$ for C1 and $V=20.51\pm 0.16$ and 
$R=19.77\pm 0.14$ for C2.
\label{fig:FlipSubIm}}
\end{center}
\end{figure}

\begin{figure}
\begin{center}
\includegraphics[scale=0.50]{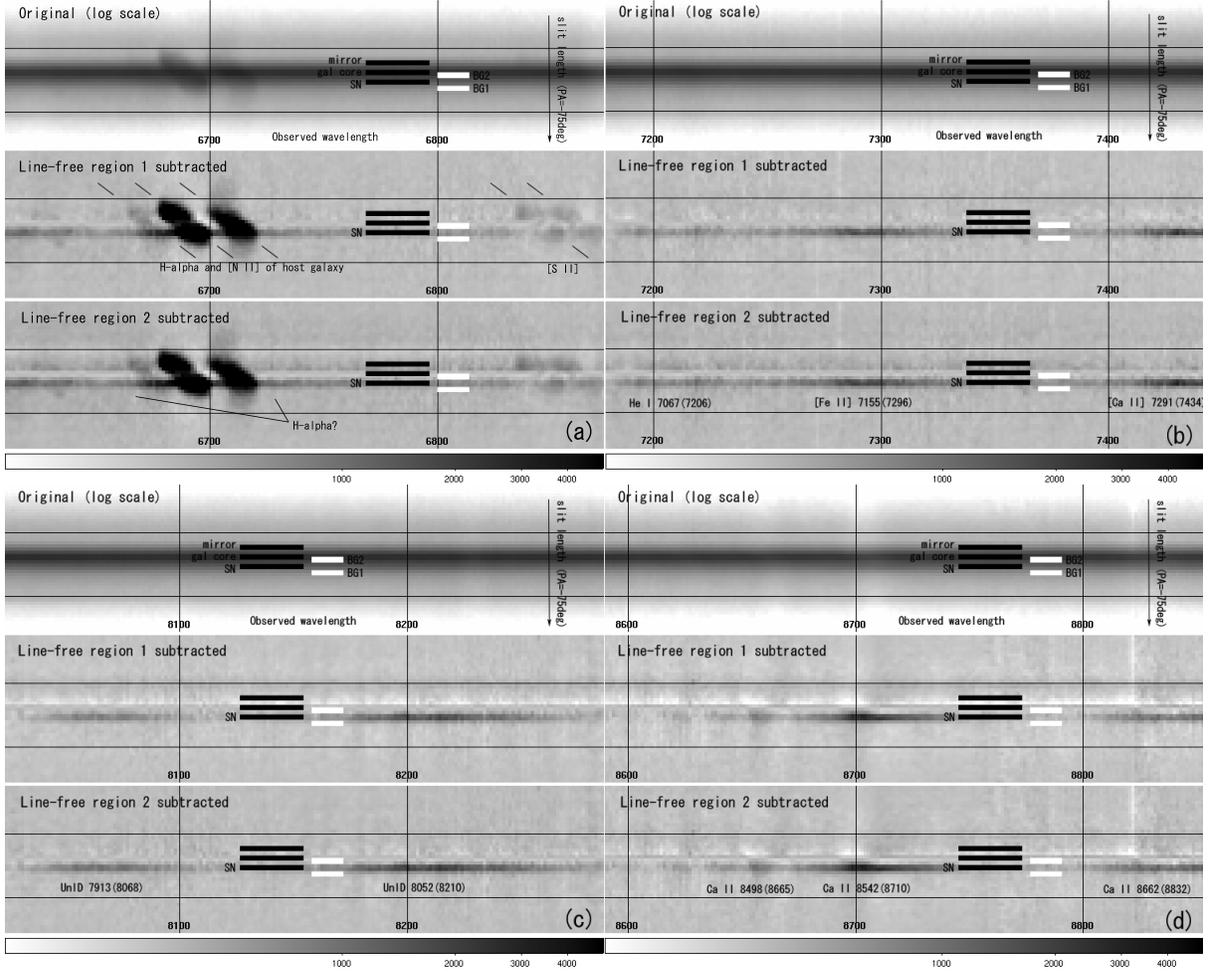}
\figcaption{
Late phase 2D spectral images around 
(a) 6750\AA\ , (b) 7300\AA\ , (c) 8150\AA\ and 8700\AA\ are shown
(observed frame).
In each wavelength region, we show (i) the original image, 
(ii) background (pattern 1) subtracted one and (iii) 
another background (pattern 2) subtracted one, from upper to lower panels.
In the original image, the sky background has already been subtracted.
The background pattern 1 and 2 are derived from the brightness
profile along the slit (i.e., vertical axis in the panel) 
at apparently line-free ($=$ pure continuum)
regions for the SN, 6983--7053\AA\ (region 1) and 
7969--8011\AA\ (region 2), respectively. Each background pattern is 
scaled by the galaxy core count at every wavelength bin (1 pixel)
and subtracted from the original 2D spectral image.
The regions used for one-dimensioning of the spectra (BG1, SN,
BG2, gal core and mirror; see Fig. \ref{fig:SPECbackg}) are 
indicated in the panel.
The width of each region is two pixels ($\simeq 0\farcs 6$).
Several broad emission lines (FWHM$> 50$\AA\ $\sim 2000$ km s$^{-1}$) 
are detected, which would be the SN origin. 
It should be noted that these images are shown to 
represent the significance of the SN spectrum even at $t=394$ days.
\label{fig:2DSpecIm}}
\end{center}
\end{figure}

For the late phase data, we should mention at first that 
the SN component was successfully detected in both imaging 
(Fig. \ref{fig:FlipSubIm}) and spectroscopic
(Fig. \ref{fig:2DSpecIm}a-d) data, 
although Smith et al. (2008a) and Agnoletto et al. (2009) reported 
their null detection in optical data between $t=364$ and $t=423.5$ days.
We assume that the better seeing condition at our observation enabled 
us to distinguish the SN component from the host galaxy nucleus.
However, it is still a difficult task to derive the intrinsic SN component
because the background galaxy has inhomogeneous structures.
For photometry on $t=$ 394 days, we assumed a symmetry for the 
brightness distribution of the host galaxy with respect to its 
polar axis (PA$\simeq -8\arcdeg$) and adopted the mirrored image 
as the background for the lack of better alternatives.
In the background-subtracted image, the SN is seen nearly 
as a point-like source embedded in a diffuse elongated component
(Fig \ref{fig:FlipSubIm}).
The PSF-photometry for the SN and nearby two stars
(C1 and C2 in Fig. \ref{fig:FlipSubIm}) provided $V=20.77\pm 0.17$ and 
$R=19.36\pm 0.14$ for the SN, 
$V=19.48\pm 0.14$ and $R=18.37\pm 0.13$ for C1 and 
$V=20.51\pm 0.16$ and $R=19.77\pm 0.14$ for C2.
Our preliminary aperture-photometry with several 
parameter sets (radii of stellar aperture and sky annulus) 
gave somewhat scattered values for the SN magnitudes,
$V=20.7\pm 0.4$ and $R=19.4\pm 0.4$, because of the 
inhomogeneous background.
To check the reliability of the uncertainty, we added an
artificial star having the same brightness at several positions 
near the SN and performed photometry of it.
The result of this test was consistent with above photometry and
we adopt $V=20.7\pm 0.4$ and $R=19.4\pm 0.4$ as the SN magnitude. 
It is noted that these magnitudes might be contaminated by 
unresolved background components, e.g., \ion{H}{ii} regions 
and/or stellar clusters in the host galaxy, and thus
they should be considered as the brighter limits.

\begin{figure}
\begin{center}
\includegraphics[scale=0.8]{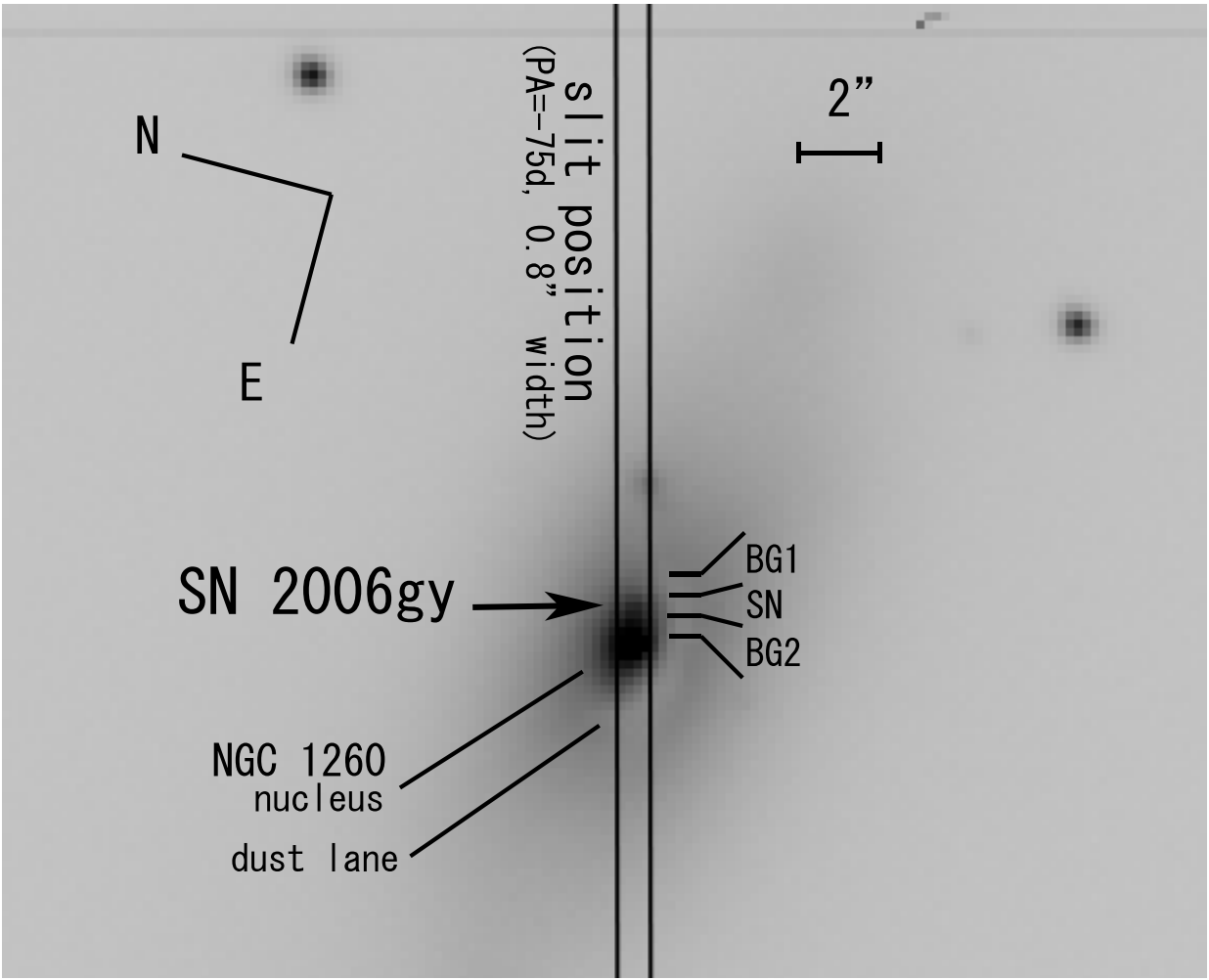}
\figcaption{
$R$ band image obtained for the slit position alignment
at the beginning of spectroscopy on 2007 Sep 18 ($t=394$ days). 
We can see SN~2006gy and the nucleus 
of the host galaxy, NGC~1260.
The SN, $\sim 0\farcs 9$ apart from the center of the galactic nucleus, 
is marked by arrow.
In the spectroscopy, the entrance slit was set as indicated by
the vertical parallel lines to lie on both the SN and 
the galactic nucleus.
We also indicate the regions which were used in one-dimensioning 
of the SN and the background (BG1, BG2) spectra (see \S 2 and \S 4).
The seeing FWHM was $0\farcs 55$ in this image.
\label{fig:SlitPos}}
\end{center}
\end{figure}

\begin{figure}
\begin{center}
\plottwo{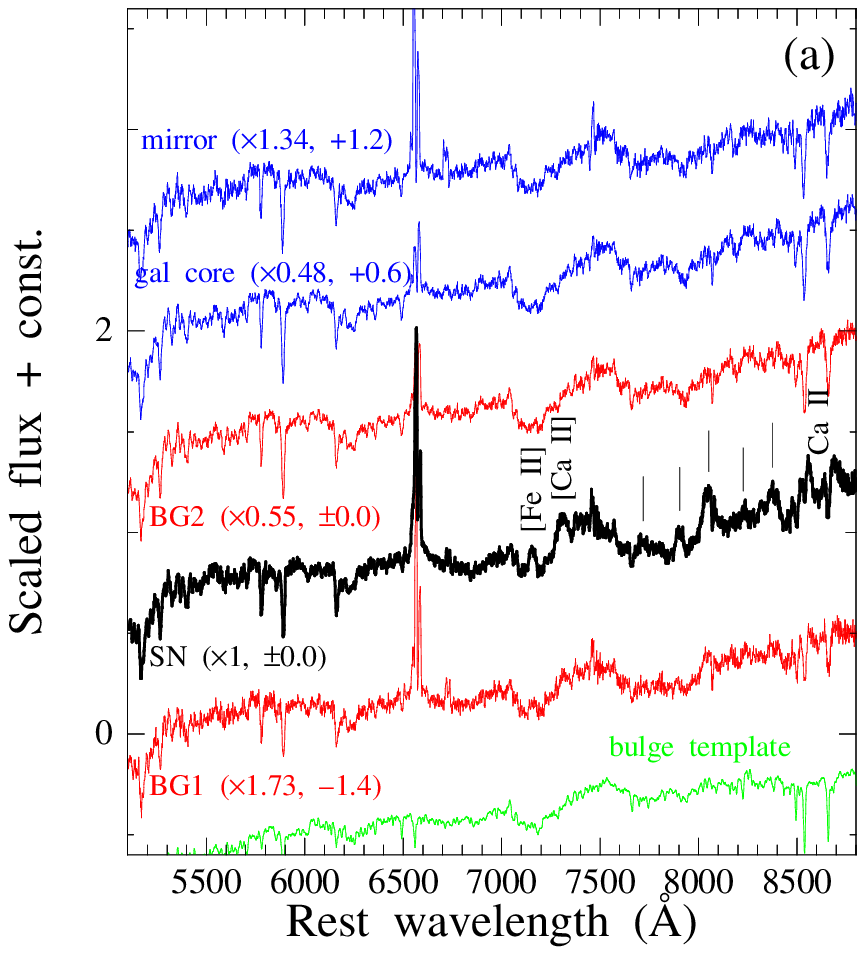}{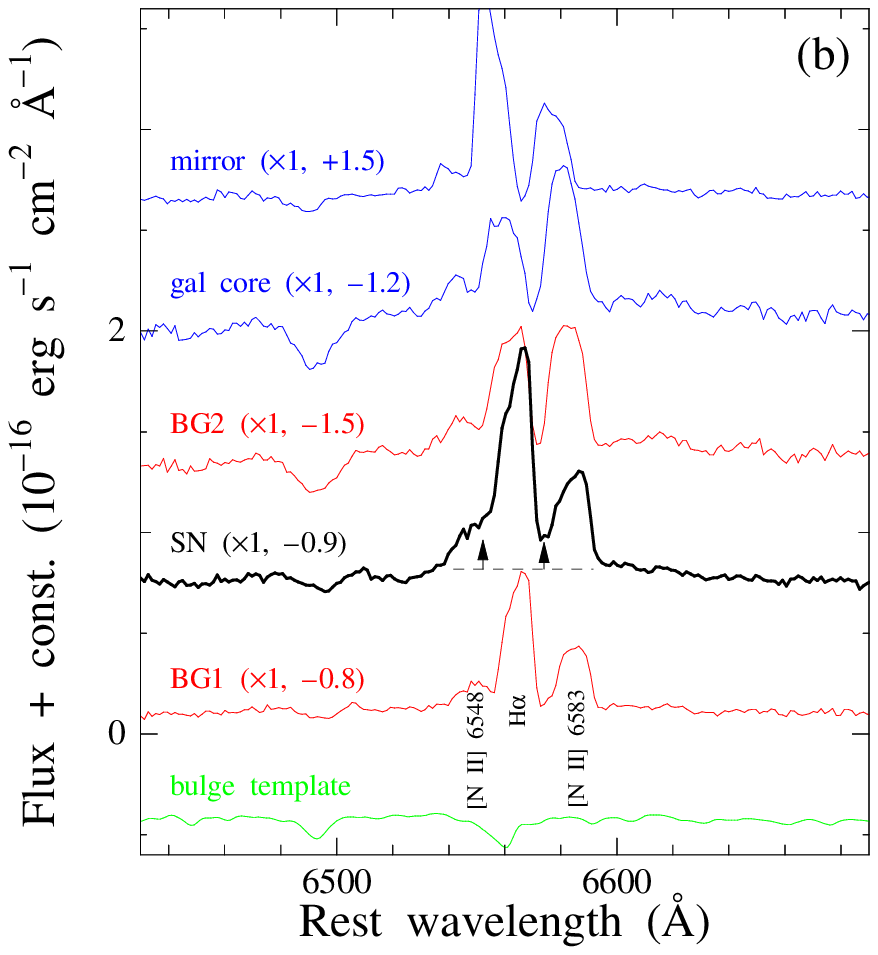}
\figcaption{
Comparison of raw spectra of the SN position and several background 
regions, BG1, BG2, galaxy core and mirror, at $t=$ 394 days:
Each spatial positions on the slit (or 2D spectrum) are shown 
in Fig. \ref{fig:SlitPos} and \ref{fig:2DSpecIm}.
The sky emission lines have been subtracted, but the 
interstellar extinction and the atmospheric absorption
have not been corrected for.
We also show the Bruzual \& Charlot (2003) model spectrum
(11 Gyr old single stellar population, solar metallicity,
$A_{V}=1$) as a template of slightly reddened galactic bulge
for comparison.
(a) Overview of the scaled spectra.
All background spectra (i.e., except for the SN one) are 
scaled to match the SN spectrum at the line-free region, 7814--7856 \AA\ 
(rest wavelength range of region 2 in Figure \ref{fig:2DSpecIm} caption).
The scaling factor and the added constant for each spectrum are 
indicated in the panel.
Some unidentified emission lines (See Table \ref{tbl:unident_line}) 
are indicated by vertical lines.
The weak 8050 \AA\ emission feature seen in the scaled BG1 spectrum 
suggests that there is a small crosstalk of the SN component
into the BG1 one.
This figure shows that (i) the continuum spectra of both
the SN and the background are mostly explained by the 
bulge component and (ii) the unidentified emission lines 
actually belong only to the SN position.
(b) Enlarged plot of original spectra around H$\alpha$ line.
Most of the line fluxes are clearly from the host galaxy,
which trace the galaxy rotation.
However, the flux excess in the SN spectrum at the 
inter-line region between H$\alpha$ and [\ion{N}{ii}] 
(indicated by arrows in the panel) imply an existence of 
intermediate-width H$\alpha$ emission line component
of $\sim 1.3\times 10^{-17}$ erg s$^{-1}$ cm$^{-2}$ \AA $^{-1}$.
\label{fig:SPECbackg}}
\end{center}
\end{figure}

In the spectroscopic data, we estimated the background component using 
the neighboring regions on both sides of the SN position, 
BG1 and BG2 along the slit (Fig. \ref{fig:SlitPos} and \ref{fig:2DSpecIm}).
Since the simple mean of BG1 and BG2 spectra clearly leads to 
over-estimation of the background component (i.e., spatial 
distribution of the background component is no longer linear along
the slit), we scaled the BG1 and BG2 spectra to match the 
SN flux at apparently line-free region, 7814--7856 \AA\ rest wavelength
before the averaging.
The scaling factors and the scaled spectra of BG1, BG2 
and some other regions are shown in Figures \ref{fig:SPECbackg}a.
Four background spectra (BG1, BG2, galaxy core 
and mirror position) resemble to one another and 
they are well explained by a slightly reddened bulge 
spectrum (Bruzual \& Charlot 2003).
From these spectra (including 2D spectral images in
Figure \ref{fig:2DSpecIm}) we recognize the existence of 
[\ion{Fe}{ii}] 7155, [\ion{Ca}{ii}] 7300 and \ion{Ca}{ii} IR
triplet emission lines in the SN spectrum.
The enlarged spectra around H$\alpha$ line (Fig. \ref{fig:SPECbackg}b)
also implies the presence of an intermediate-width 
(15--30\AA\ $\simeq$ 700--1400 \kms ) H$\alpha$ emission line.
They are discussed later.
Finally, we adopted the mean of the scaled BG1 and BG2 spectra
as the background component, and then subtracted it from the SN 
position spectrum to derive the pure SN spectrum.
This method leads to complete elimination of the continuum 
component of the SN, at least, around the line-free regions.
The resultant SN spectrum provides the faint-end limit 
of the magnitude of the SN as $V\lesssim 21.7$ and 
$R\lesssim 21.1$ (as far as we ignore the contamination 
by the possible unresolved component as mentioned above), 
which are consistent with the results of our photometry.
It is noted that there is a non-negligible difference between 
the normalized BG1 and BG2 spectra, especially at bluer wavelengths, 
which may lead to a large uncertainty for the resultant SN flux
at $t=394$ days up to $\pm 60$\% (Fig. \ref{fig:SPECobs2}).
This uncertainty is larger than the error of estimated extinction 
toward the SN (see \S\ref{sec:spec}).
In this paper we mainly discuss the emission line components 
at longer than 5800 \AA\ for the late phase spectrum and 
therefore this uncertainty does not affect our discussion.

\section{Light Curve Analysis}
\label{sec:LC}

\begin{figure}
\begin{center}
\plottwo{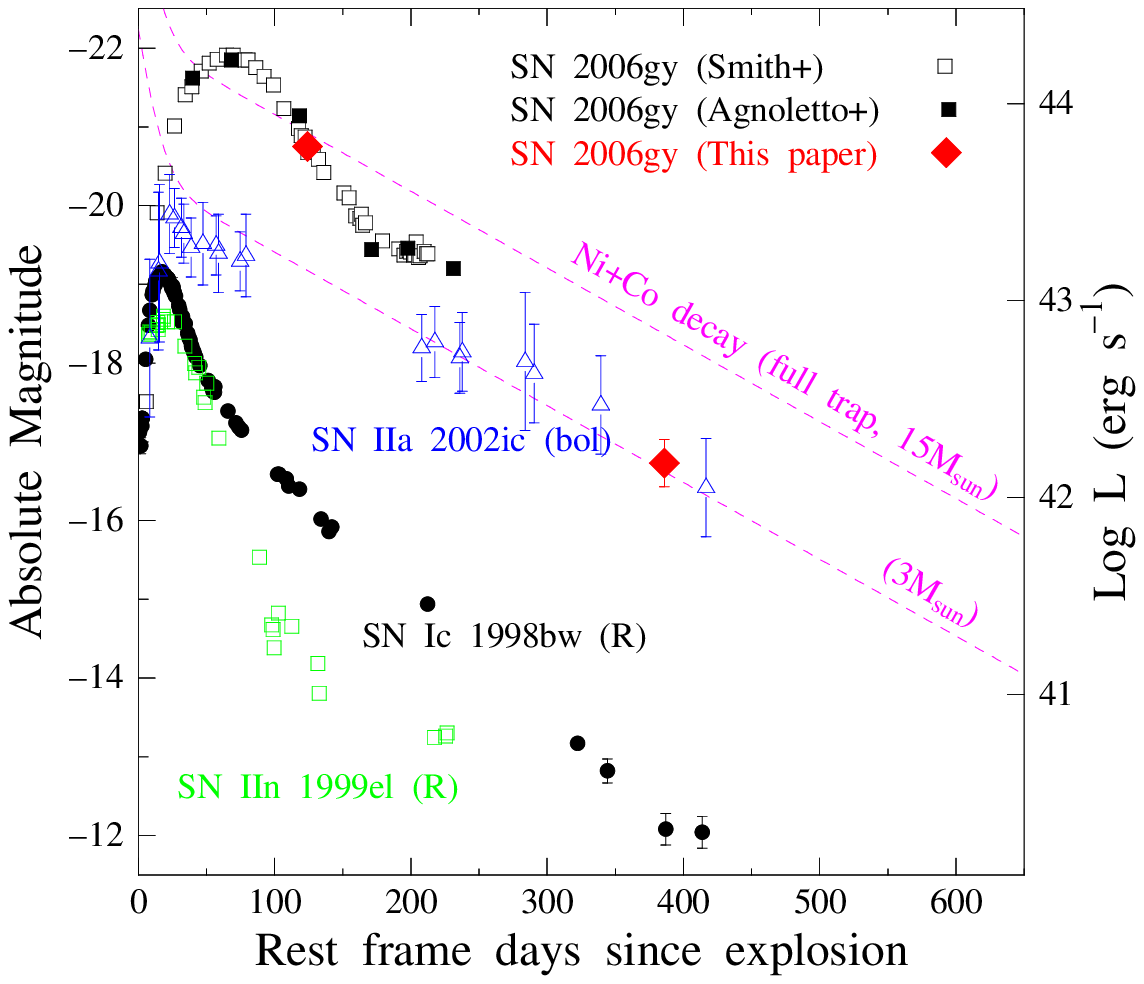}{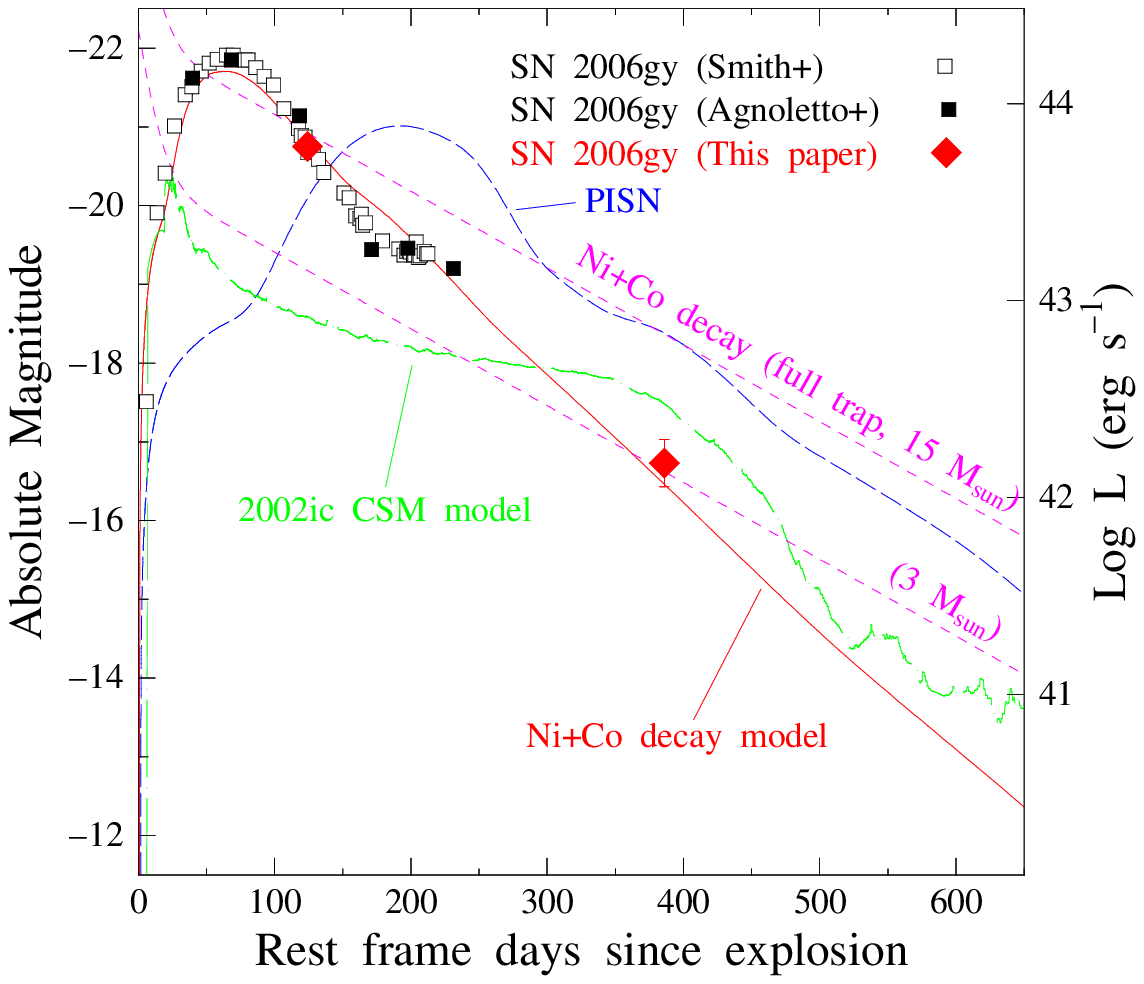}
\figcaption{
(Left) Absolute $R$ band light curve of SN 2006gy compared with 
various types of SNe.
For SN 2006gy, $\mu = 34.45$ and $A_R = 1.68$ are used.
The right vertical axis is the luminosity assuming zero bolometric correction.
Open and filled black squares are $R$ band magnitudes of SN 2006gy 
by Smith et al. (2007) and Agnoletto et al. (2009), respectively.
Filled red diamonds denote the $R$ magnitude
estimated from our Subaru observations.
Bolometric magnitudes of SN IIa 2002ic (Deng et al. 2004),
R band magnitudes of SN Ic 1998bw (Patat et al. 2001) 
and SN IIn 1999el (Di Carlo et al. 2002) are shown with
blue open triangles, black filled circles, and green open squares, 
respectively.
The dashed magenta lines show decay energy from $15 \Msun$
and $3 \Msun$ of \Nifs +\Cofs ,
respectively (decay data are from Nadyozhin 1994).
(Right) SN 2006gy LC compared with some LC models.
For LC of SN 2006gy and the dashed magenta lines are same as
the left panel.
Red solid line shows a synthetic LC of radioactive decay model
by Nomoto et al. (2007).
The parameters are ($\KE/10^{51} {\rm ergs}, \Mej/\Msun, \Mni/\Msun$) 
= ($64$, $53$, $15$).
This model roughly follows the overall LC from the explosion 
to $t=394$ days, although the tail of the observed LC around 
$t=200$ days is flatter than expected by this model.
Blue dashed line denotes a model with 
($\KE/10^{51} {\rm ergs}, \Mej/\Msun, \Mni/\Msun$)
= ($65$, $166$, $15$) imitating a pair-instability SN (PISN).
Green dashed line is a CSM interaction model for SN 2002ic
(Nomoto et al. 2005).
\label{fig:LCabs}}
\end{center}
\end{figure}

Figure \ref{fig:LCabs} shows the $R$ band light curve (LC) of 
SN 2006gy constructed from the published data
(Smith et al. 2007; Agnoletto et al. 2009) and our Subaru ones. 
It is compared with bright Type Ic SN 1998bw (Patat et al. 2001)
and Type IIa SN 2002ic (Deng et al. 2004)
and Type IIn SN 1999el (Di Carlo et al. 2002).
There exists a large variety among the LCs of Type IIa and IIn SNe.
The LC of SN IIa 2002ic declines very slowly.\footnote{We assume
that the explosion date is 2002 Nov 3 \citep{ham03} although it
is unclear and an exact LC comparison is not easy.}
The slope is flatter than that the LC powered by the 
\Cofs\ $\rightarrow$ \Fefs\ decay with fully trapped $\gamma$-rays.
This behavior is also seen in other SNe, \eg SN IIa 1997cy and SN IIn 1988Z  
(\eg Germany et al. 2000; Turatto et al. 1993; Turatto et al. 2000).
On the other hand, the LC of SN 1999el declines faster 
than SNe 2002ic and 1998bw. 
Similar behavior is also seen in the LC of Type IIn SN 1998S
(Fassia et al. 2000).
In another Type IIn SN 1994W, a sudden decline of the LC was observed 
after $\sim 100$ days plateau
(Sollerman, Cumming \& Lundqvist 1998, Chugai et al. 2004).
The main heating source of Type IIn and IIa SNe 
is thought to be the interaction between 
SN ejecta and CSM (\eg Turatto et al. 1993; Chugai et al. 2004),
and this heterogeneity in the LC may reflect the structure,  
the mass and the kinetic energy of the SN ejecta as well as 
the structure (density and its slope) of CSM.

In contrast to Type IIn SNe, the LCs of SNe where \Nifs\ is 
the dominant heating source show similar shapes.
The LC declines faster than the \Cofs\ decay line because 
a fraction of $\gamma$-rays escape without interacting with the SN ejecta.
The deviation from the \Cofs\ decay line gradually becomes 
larger because the expanding SN ejecta gets
thinner against $\gamma$-rays.
The LC of Type Ic SN 1998bw is fully explained by this scenario 
(Patat et al. 2001; Maeda et al. 2003).

The early LC of SN 2006gy tends to be leveled around $t=200$ days.
Our Subaru data shows that the LC fades by $\sim 3$ mag 
from $t \sim 200$ to $\sim 400$ days, which is consistent with 
the results of NIR photometry at $t=405$ and 468 days
(Smith et al. 2008a).
The decline rate during this period is clearly faster 
than that of SN 2002ic and the \Cofs\ decay line,
while it seems comparable to that of SN 1998bw.
In other words, the LC of SN 2006gy does not show any particularly 
peculiar behavior that is not explained by the \Cofs\ decay.
However, we cannot exclude the possibility that the LC 
is almost leveled at $t \sim 200-350$ days and suddenly declines
just before $t \sim 394$ days.

In the right panel of Figure \ref{fig:LCabs}, the \Nifs $+$ \Cofs\ decay 
model presented by Nomoto et al. (2007) is also shown.
They computed a bolometric LC and showed that the model with 
$\Mej = 53 \Msun$, $\KE=64 \times 10^{51}$ erg and $\Mni = 15 \Msun$ 
explains the observed LC up to $t \sim 200$ days.
Here $\Mej$, $\KE$ and $\Mni$ are the mass of the ejecta, 
the kinetic energy of the ejecta, and the mass of \Nifs\ 
ejected by the explosion, respectively.
The model LC is in reasonably good agreement with 
our late phase observation.

If we assume that the radioactive decay is the main energy
source of emission at $t=394$ days, $\Mni \gtrsim 3 \Msun$ is 
needed to explain the $R$ band flux.
Although the model LC for SN IIa 2002ic (Nomoto et al. 2005)
is not consistent with that of SN~2006gy, we cannot preclude 
the CSM interaction model because of the large diversity of
LCs in Type IIn/IIa SNe.
We note that the model by Woosley et al. (2007) is also 
consistent with our observation.
However, the diffusion model by Smith \& McCray (2007), 
predicting a fast decline after the maximum 
($\sim -13$ mag at $t = 300$ days), is inconsistent with our data.
Also, the LCs of the PISN model (blue dashed line) is not
consistent with the observed LC.

\section{Spectral Analysis}
\label{sec:spec}

\begin{figure}
\begin{center}
\includegraphics[scale=1.0]{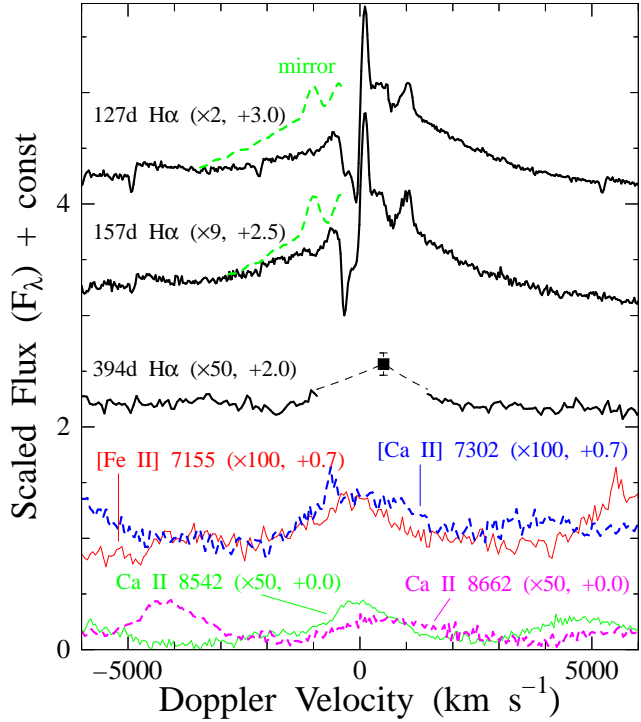}
\figcaption{
The line profiles of H$\alpha$ of our three epoch 
spectra (black) plotted against Doppler velocity.
For the spectrum at $t=394$ days, the real profile
of the  H$\alpha$ and [\ion{N}{ii}] lines is hardly to be 
known by the contamination of the background 
\ion{H}{ii} region components (see \S\ref{sec:backsub}).
Alternatively, we plot the flux at 6574 \AA\ which is 
a somewhat reliable estimate from the residual flux at 
the trough between H$\alpha$ and [\ion{N}{ii}] 6584 lines.
The error bar denotes the uncertainty of its continuum level 
(see Fig. 5).
The green dashed line labeled `mirror' is the reflected red-side 
profile across $v=0$ \kms (cf. Fig. 5 in Smith et al. 2007).
The lower four lines are the Fe and Ca lines at $t=394$ days 
({\it solid red}: [\ion{Fe}{ii}] 7155, 
{\it dashed blue}: [\ion{Ca}{ii}] 7302 (average of 7291 and 7323), 
{\it solid green}: \ion{Ca}{ii} 8542, 
{\it dashed magenta}: \ion{Ca}{ii} 8662). 
The scaling factor and the added constant in each spectrum are
indicated in the panel.
\label{fig:SPECvel}}
\end{center}
\end{figure}

The spectra of SN 2006gy are shown in Figure \ref{fig:SPECobs},
and closed-up profiles of some lines are shown in Figure \ref{fig:SPECvel}.
The redshift $z=0.01967$, derived from the diffuse H$\alpha$ emission 
line of the host galaxy (cf. $z=0.01910$ derived at the nucleus), has been 
corrected for.
We assumed the total extinction of $E(B-V)=0.628\pm 0.094$ (or
$A_{R}=1.68\pm 0.25$) toward the SN, which is the sum of the 
extinction in the Milky Way [$E(B-V)=0.16$; Schlegel, 
Finkbeiner, \& Davis 1998] and that within the host galaxy 
[$E(B-V)=0.468\pm 0.094$, derived from $A_{R}=1.25\pm 0.25$ and
$A_{\mbox{\scriptsize Landolt }R}=2.673\times E(B-V)$; Smith et al. 2007,
Schlegel et al. 1998].
The equivalent width (EW) of the Galactic 
\ion{Na}{i} D$_{1}$D$_{2}$ absorption lines in our high 
resolution spectrum (EW$=1.2$ \AA ) is almost
consistent with $E(B-V)=0.16$ using an empirical
relation of $E(B-V)=-0.01 + 0.16\times EW$(\AA ) suggested by Turatto, 
Benetti, \& Cappellaro (2003).
Although the EW of the host galaxy \ion{Na}{i} D line 
in the same spectrum suggests a larger extinction within the
host galaxy ($E(B-V)\sim 1$), it should be partly circumstellar
origin because some of other spectral features (H$\alpha$ and 
\ion{Fe}{ii}) are accompanied with narrow absorption components
in their P-Cyg type profiles.
Our late time spectrum suffers from large uncertainty 
in the EW of \ion{Na}{i} D line because of
the faint SN spectrum and the uncertain continuum level
(see \S\ref{sec:backsub}).

Like other Type-IIn SNe, our early phase spectra of SN 2006gy 
are characterized by a strong H$\alpha$ emission line as well as 
many \ion{Fe}{ii}, \ion{He}{i} and \ion{Na}{i} lines.
As shown in Figure \ref{fig:SPECvel}, the H$\alpha$ emission line
in the spectra at $t=127$ days has a narrow P-Cyg type component 
near the peak, of which the marginally resolved 
emission line has a width of $\sim 120$ \kms\ FWHM, 
and a broad wing component exceeding $\pm 5000$ \kms.
The broad component is clearly asymmetric, being brighter 
in the red wing than in the blue.
Smith et al. (2007) took a high-resolution spectrum 
at $t=96$ days ($R\approx 4500$) and suggested that 
the asymmetric profile is
produced by broad absorption component at the blue side.
This component had a sharp blue edge at $\sim -4000$ \kms, 
which is considered to be an effective speed of the 
SN blast wave (Smith et al. 2007).
This asymmetric profile was seen through $t=157$ days, although 
the absorption component became less significant.
For some features, e.g., \ion{Fe}{ii} 5535 and \ion{Na}{i} D
at $t=127$ days, broad blue-shifted absorption components up to
$4000-5000$ \kms\ are also significant (Fig. \ref{fig:SPECobs}).
The narrow absorption component of the H$\alpha$ peak
considerably evolved from $t=96$ to 157 days.
It splits into possible double components by 127 days, and the whole
width between both edges of the absorption feature increased
from $260$ \kms\ to $690$ \kms\ .
These suggest an ongoing interaction between the ejecta
and the dense CSM.

\begin{figure}
\begin{center}
\includegraphics[scale=.8]{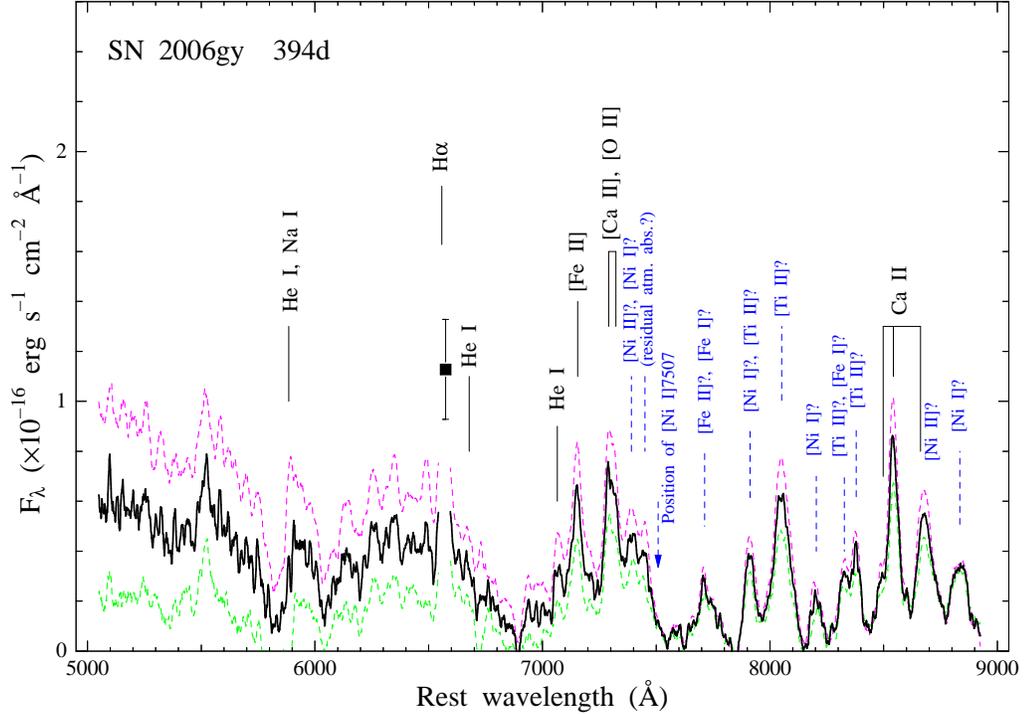}
\figcaption{
Our low resolution spectrum of SN 2006gy at $ t=394$ days,
which are smoothed with neighboring eight pixels
(corresponding to the width of the slit).
The interstellar extinction within our galaxy and NGC~1260 
have been corrected for.
The atmospheric absorption bands are eliminated using the spectrum 
of spectrophotometric standard stars obtained on the same night.
We derived this spectrum by subtraction of the average of the
scaled BG1 and BG2 spectra from the SN position one (Fig. 
\ref{fig:SPECbackg}).
The upper dashed line denotes the SN spectrum in which only 
the scaled BG2 is subtracted (see \S\ref{sec:backsub}).
On the other hand, the lower dashed line denotes that in 
which only the scaled BG1 is subtracted.
We consider that these lines may represent the uncertainty 
of the derived SN spectrum.
In this plot we removed the spectrum near the H$\alpha$ line
because its profile is distorted by the subtraction
of the background \ion{H}{ii} region component (see \S\ref{sec:backsub}).
Alternatively, we plotted the possible intrinsic component at H$\alpha$
emission line, estimated from the residual flux at the 
trough between H$\alpha$ and [\ion{N}{ii}] 6584 lines 
(Fig. \ref{fig:SPECbackg}).
There are several identified and unidentified emission
lines at 7000--8800 \AA ,
which are listed in Table \ref{tbl:unident_line}.
The position of [\ion{Ni}{i}] 7507, which should be as strong as
[\ion{Ni}{i}] 7908, is indicated by an arrow (see \S 5.3). 
\label{fig:SPECobs2}}
\end{center}
\end{figure}

\begin{figure*}
\begin{center}
\includegraphics[scale=0.9]{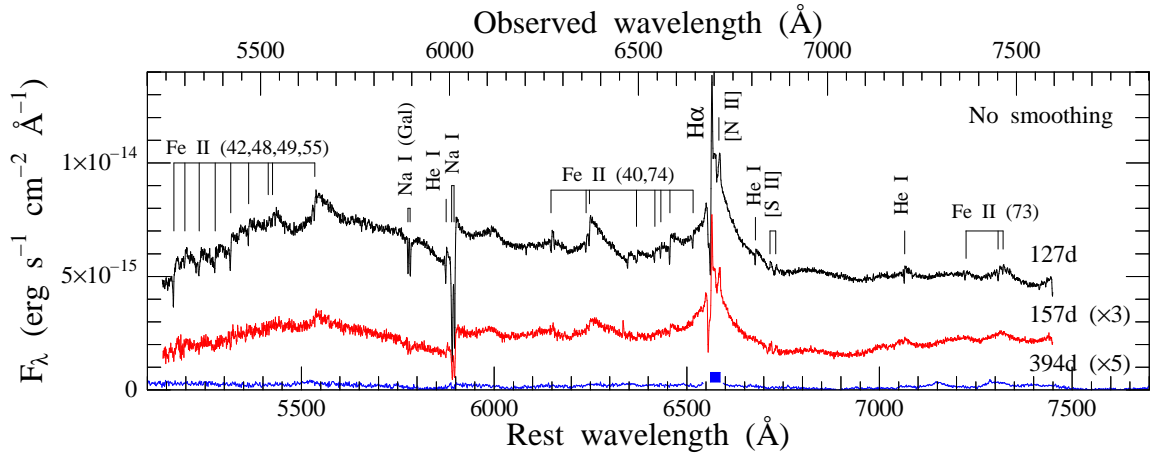}
\figcaption{
High resolution spectra of SN 2006gy at $t=$ 127 and 157 days are shown
together with the low resolution spectrum at 394 days.
The interstellar extinction and atmospheric absorption bands have
been corrected for as in Figure \ref{fig:SPECobs2}.
The scaling factor of each spectrum is indicated in the panel.
In the earlier spectra, \ion{Fe}{ii}, \ion{Na}{i}, \ion{He}{i} 
and H$\alpha$ lines show pseudo P Cyg-like profiles with steep slope 
between the blue-shifted absorption and the red-shifted emission components.
Most of the absorption components are much narrower than the
accompanied emission lines.
\label{fig:SPECobs}}
\end{center}
\end{figure*}

In the late phase spectrum (Fig. \ref{fig:SPECobs2}), 
we can see several emission lines, mainly in red wavelengths, including
[\ion{Ca}{ii}] 7291, 7323 (7302 in average), \ion{Ca}{ii} IR triplet and 
[\ion{Fe}{ii}] 7155, which are listed in Table \ref{tbl:unident_line}.
On the other hand, the bluer spectrum is likely to be dominated by 
continuum component and/or heavily blended emission lines.
A broad absorption trough (up to $-5000$ \kms ) possibly belonging 
to \ion{Na}{i} D is significant in this region. 
Additionally, as mentioned in \S \ref{sec:backsub}, there is a possible
H$\alpha$ emission line having an intermediate-width (700--1400 \kms ).
If the flux excess at the minimum point (6574 \AA ) between 
H$\alpha$ and [\ion{N}{ii}] 6584 lines is actually due to the 
intermediate-width H$\alpha$ component (Fig. \ref{fig:SPECbackg}b), 
its observed flux is derived to be 
$\simeq 6 \times 10^{-17}$ erg s$^{-1}$ 
cm$^{-2}$ \AA$^{-1}$ above the continuum level of
$\simeq  5\times 10^{-17}$ erg s$^{-1}$ cm$^{-2}$ \AA$^{-1}$.
This continuum flux is consistent with the upper-limit 
($\lesssim 2\times 10^{-17}$ and $\lesssim 9\times 10^{-17}$
erg s$^{-1}$ cm$^{-2}$ \AA$^{-1}$ before and after the 
extinction-correction, respectively) set by Smith et al. (2008a).
Assuming that the flux excess is actually due to an H$\alpha$ line 
of which the line width is $\sim 20$ \AA , we can derive the luminosity 
of the line flux as $\sim 1\times 10^{39}$ erg s$^{-1}$.

\begin{figure}
\begin{center}
\includegraphics[scale=0.8]{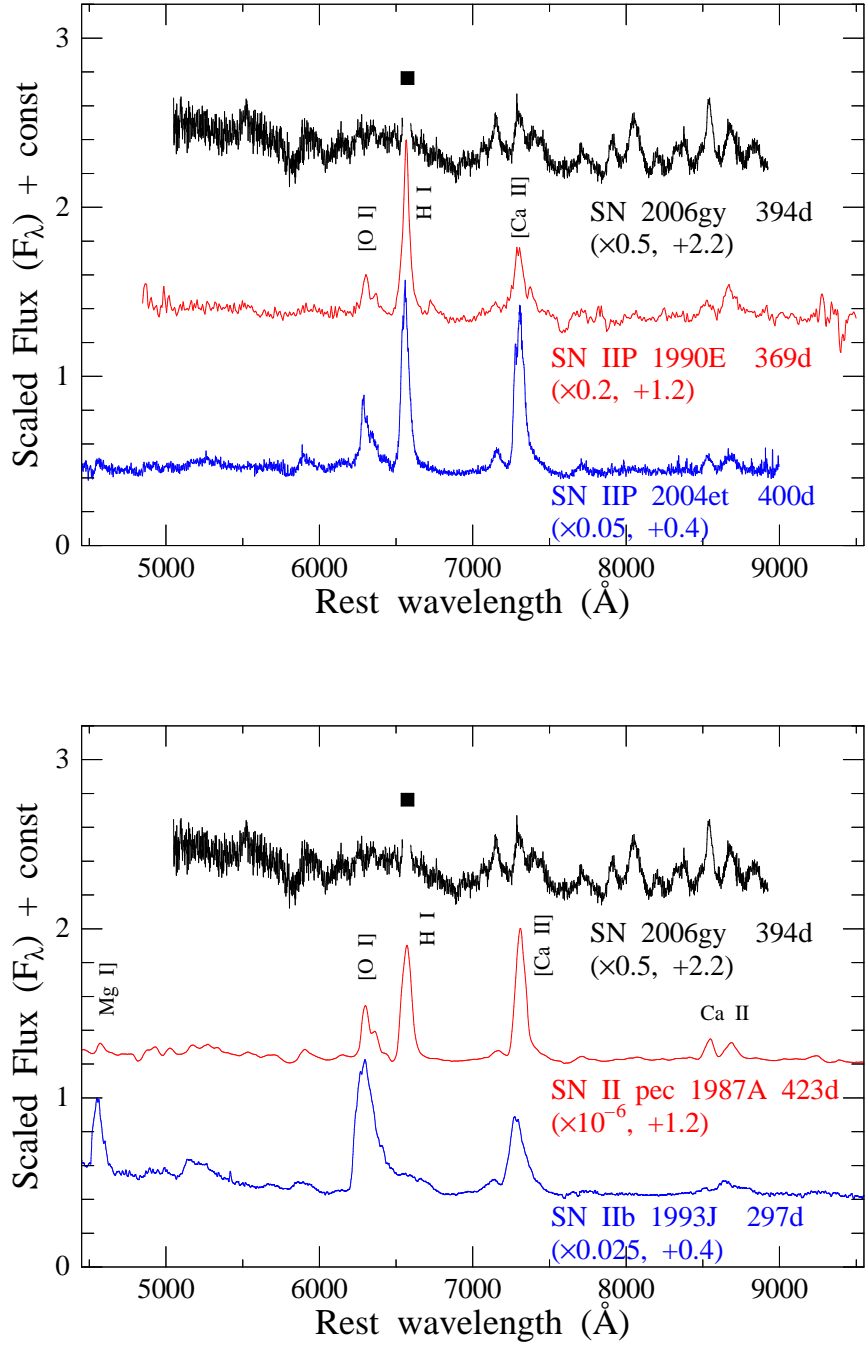}
\figcaption{
Comparison of late time spectra with (upper panel) SNe IIP
1990E (G\'omez \& L\'opez 2000) and 2004et (Sahu et al. 2006)
and, (lower) peculiar SN II 1987A (Pun et al. 1995) 
and SN IIb 1993J (Barbon et al. 1995).
The unit of the vertical axis is $10^{-16}$ erg s$^{-1}$ cm$^{-2}$
\AA $^{-1}$.
Each flux is scaled and shifted for convenience of comparison
and those values are indicated in the panel. 
The epoch is given as days from estimated explosion date ($t$).
\label{fig:SPECcomp1}}
\end{center}
\end{figure}

\begin{figure}
\begin{center}
\includegraphics[scale=0.8]{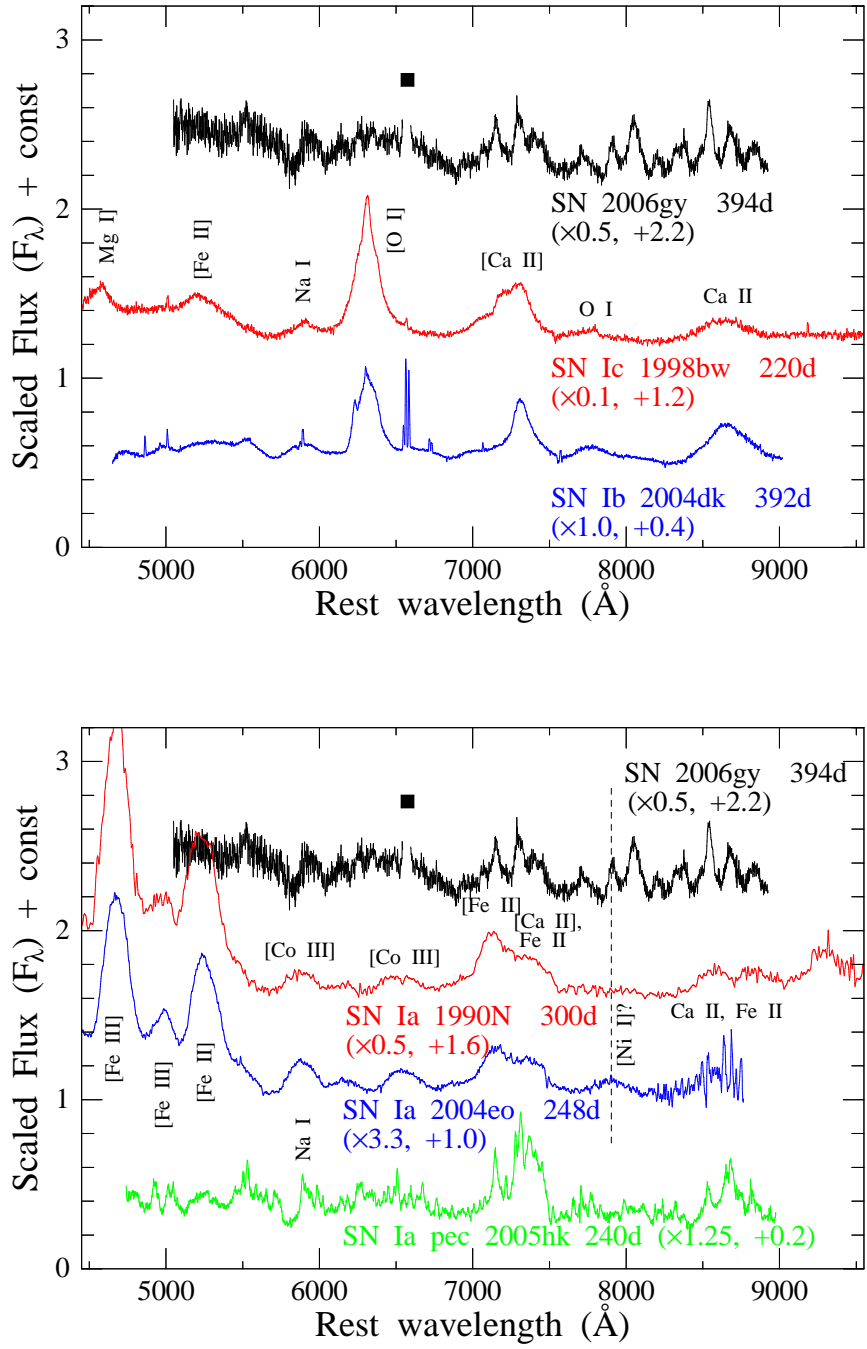}
\figcaption{
Comparison with (upper panel) 
SN Ic 1998bw (Patat et al. 2001) and SN Ib 2004dk
(Maeda et al. 2008), and (lower) 
SNe Ia 1990N (G\'omez \& L\'opez 1998),
2004eo (Pastorello et al. 2007) and peculiar SN Ia
2005hk (Sahu et al. 2008).
\label{fig:SPECcomp3}}
\end{center}
\end{figure}

In Figures \ref{fig:SPECcomp1}--\ref{fig:SPECcomp3}, the late time 
spectrum of SN 2006gy is compared with those of various classes of SNe.
Unlike other slowly-declining Type IIn/IIa SNe, the H$\alpha$ 
emission line is neither strong nor wide.
The emission of \ion{Ca}{ii} IR triplet in SN 2006gy is 
much narrower than those in other Type IIn/IIa SNe, 
suggestive of a slow SN ejecta.
The presence of forbidden lines [\ion{Fe}{ii}] 7155 and 
[\ion{Ca}{ii}] 7302 is also peculiar
for a Type IIn/IIa SN.
On the other hand, this SN does not show strong [\ion{O}{i}] 6300, 6364 
emission line, which is generally strong in Type II and Ib/c SNe 
having forbidden [\ion{Ca}{ii}] 7302 line (Fig. \ref{fig:SPECcomp1}
and \ref{fig:SPECcomp3})\footnote{The [\ion{O}{i}] line is also 
absent in Type Ia SNe because of the low oxygen abundance 
in the ejecta (see \eg Kozma et al. 2005)}.
Some of the intermediate-width emission lines, marked with dashed 
lines in Figure \ref{fig:SPECobs2}, are unusual for any type of SNe. 
The presence of the wide absorption trough at \ion{Na}{i} D 
is also unusual for Type IIn/IIa SNe.
Similar feature is seen in the late phase spectrum of a peculiar 
Type Ia SN 2005hk (Fig. \ref{fig:SPECcomp3}).
We discuss these peculiar properties of SN 2006gy in the next section.

\section{Discussion}
\label{sec:discussion}

\subsection{CSM Interaction or Radioactive Decay Heating?}
\label{subsec:powersource}

The luminosity of SN 2006gy at the late phase is consistent with the 
radioactive decay model as shown in Figure \ref{fig:LCabs}.
The LCs of Type IIn SNe show a large variety and
some of which might also be in fair agreement with SN 2006gy.
Thus, it seems difficult to discriminate the main heating source
of SN 2006gy only from the LC.
The late time spectrum gives some indications on this issue.

\begin{figure}
\begin{center}
\includegraphics[scale=0.8]{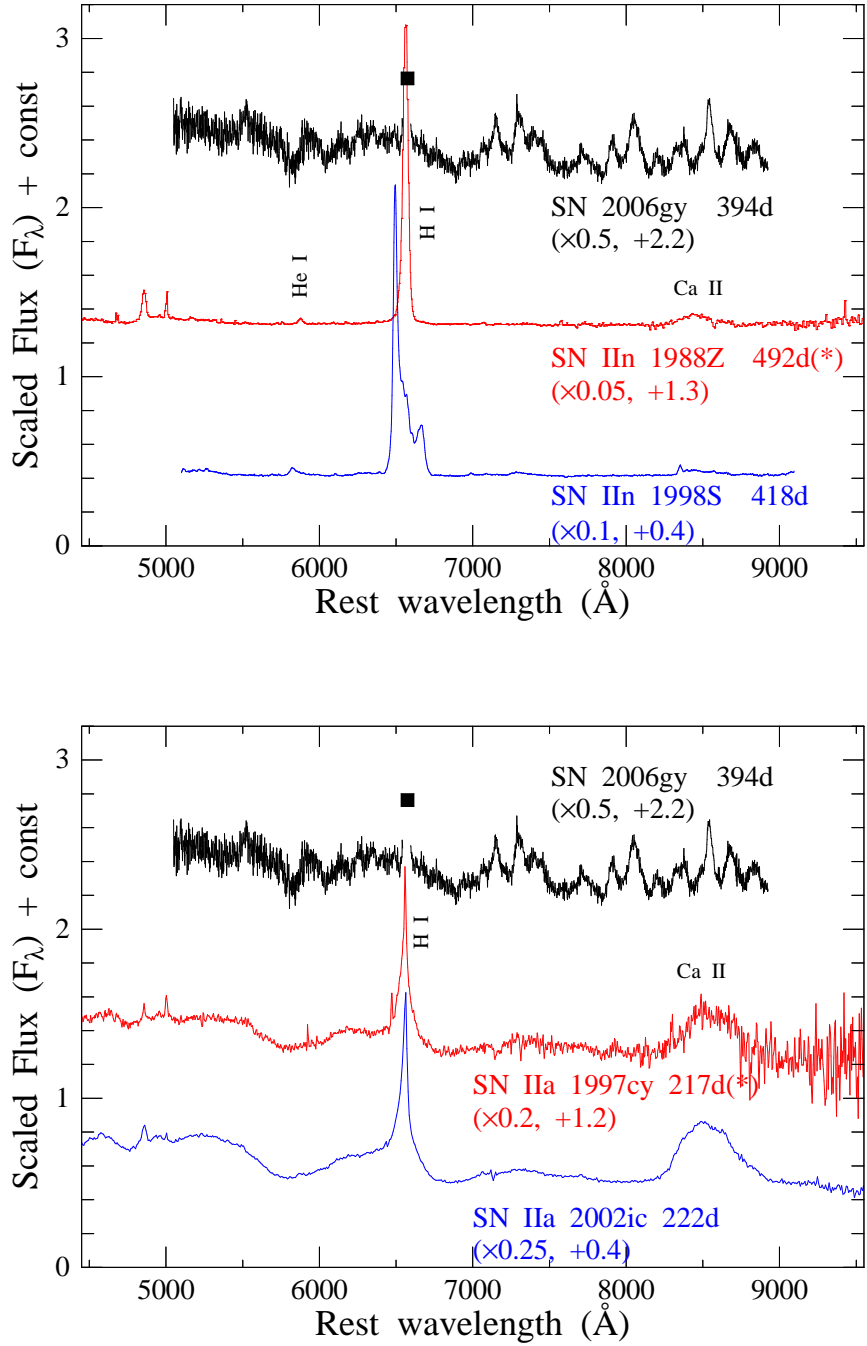}
\figcaption{
Comparison with (upper panel) 
SNe IIn 1988Z (Turatto et al. 1993) and 1998S
(Pozzo et al. 2004), and (lower) SNe IIa 1997cy 
(Germany et al. 2000; Turatto et al. 2000) and 2002ic
(Deng et al. 2004).
The epoch is same as in Figure \ref{fig:SPECcomp1}, 
except for SNe 1988Z and 1997cy
(days since the discovery, marked with $*$).
The H$\alpha$ emission line of SN 1998S has a double peak profile
with a wide separation.
\label{fig:SPECcomp2}}
\end{center}
\end{figure}

SN 2006gy is classified as Type IIn from earlier
spectra showing strong H$\alpha$ emission line.
However, as mentioned in \S \ref{sec:spec}, the late time spectrum 
is quite atypical for Type IIn/IIa SNe (Fig. \ref{fig:SPECcomp2}).
The emission lines of \ion{Ca}{ii} IR triplet are narrow and not
blended. 
There are several forbidden lines, e.g., [\ion{Fe}{ii}] 7155 
and [\ion{Ca}{ii}] 7302.
The luminosity of the possible H$\alpha$ emission 
line at 394 days ($\sim 1\times 10^{39}$ erg s$^{-1}$) is considerably
smaller than those of Type IIn SNe 1988Z ($2.0\times 10^{41}$
erg s$^{-1}$ at 492 days; Turatto et al. 1993), 1995Z
($3.2\times 10^{40}$ erg s$^{-1}$ at 716 days; Fransson et al. 2002)
and 2006tf ($7.9\times 10^{40}$ erg s$^{-1}$ at 445 days after 
discovery; Smith et al. 2008b) and also Type IIa SN 2002ic 
at earlier phase [$(2-3)\times 10^{41}$ erg s$^{-1}$ at 222 days; 
Deng et al. 2004].
It is rather consistent with the pure radioactive decay model of 
a typical type II SN ($\sim 0.1$ M$_{\odot}$ \Cofs\ and 
$\sim 10$ M$_{\odot}$ envelope; Chugai 1991).
Thus, the observed H$\alpha$ emission flux does not 
require any Type IIn-like CSM interaction at $t=394$ days.
Therefore, the strong CSM interaction expected from the conspicuous 
H$\alpha$ emission and the flattened LC before $\sim$ 200 days seems 
to have faded by 394 days.
If the observed $R$ band luminosity is purely originated from
radioactive decay, $\Mni\gtrsim 3\Msun$ is necessary as discussed
in \S\ref{sec:LC}.

However, the pure radioactive decay model also seems to 
have an inconsistency.
Core-collapse SNe generally show the [\ion{O}{i}] 6300, 6364 
line at late time except for Type IIn/IIa SNe,
while SN 2006gy showed little or no [\ion{O}{i}] emission
(Figs. \ref{fig:SPECcomp1}--\ref{fig:SPECcomp3}).
As for Fe lines, in Type Ic SN 1998bw, a blend of [\ion{Fe}{ii}] lines 
is seen around 5200 \AA\ and the [\ion{Fe}{ii}] 7155 line has a dominant 
contribution to the emission line around 7200 \AA,
while only [\ion{Fe}{ii}] 7155 line is seen in SN 2006gy.
The LC of SN 1998bw is fully explained by the heating of 
$\sim 0.4 \Msun$ of \Nifs\
(Iwamoto et al. 1998; Nakamura et al. 2001; Maeda et al. 2006).
If the peak luminosity of SN 2006gy is powered by the decay of 
\Nifs, the required mass of \Nifs\ is $\gsim 10\Msun$, 
more than 20 times larger than that of SN 1998bw.
This would result in stronger Fe lines in SN 2006gy.
The same discussion holds true in comparison with Type Ia 
SNe, which eject $\sim 0.6 \Msun$ of \Nifs\ on average
(e.g., Stritzinger et al. 2006).
Both SNe 1990N and 2004eo showed strong and broad [\ion{Fe}{ii}] 
features around 5200 \AA\ and 7150 \AA\ (Fig. \ref{fig:SPECcomp3}).
Thus, the weakness of [\ion{O}{i}] and [\ion{Fe}{ii}] lines 
seem against the \Nifs\ heating scenario.
However, this peculiarity in the line strength might be
explained by a moderately high density of the 
line emitting region.
We will discuss it in \S \ref{subsec:density}.

\subsection{Moderately High Density of Line Emitting Region}
\label{subsec:density}

The presence of the forbidden [\ion{Fe}{ii}] 7155 and [\ion{Ca}{ii}] 7302
lines suggests the density of the emitting region in SN 2006gy 
is less than those in other Type IIn/IIa SNe.
However, it is likely that the density is still higher than those of 
other type SNe; the line flux ratio
$F(7302)/F({\rm IR\ triplet})$ of SN 2006gy is only $\simeq 0.5$
(Table \ref{tbl:unident_line}), while
$F(7302)/F({\rm IR\ triplet}) > 1$ in SNe of other types 
(Fig. \ref{fig:SPECcomp1} and \ref{fig:SPECcomp3}; see also 
G\'omez \& L\'opez 2000).

The value of $F(7302)/F({\rm IR\ triplet})\simeq 0.5$ suggests 
an electron density of the emitting region 
$N_{e}\simeq 10^{8}$--$10^{9}$ cm$^{-3}$ (Ferland \& Persson 1989; 
Fransson \& Chevalier 1989).
This is consistent with the existence of the [\ion{Fe}{ii}] 7155
because its critical density of electron is not so small 
(a few times $10^{8}$ cm$^{-3}$; Zickgraf 2003).
In such a high density region, the forbidden line [\ion{O}{i}] 
6300, 6364 would be effectively quenched, because its critical 
density is only a few times $10^{6}$ cm$^{-3}$.
Therefore, the low $F(7302)/F({\rm IR\ triplet})$ value and
the absence of [\ion{O}{i}] 6300, 6364 in SN 2006gy can be qualitatively
explained by a moderately high electron density of the emitting 
region, which would be less than those of Type IIn/IIa SNe and 
more than other Type II and Ib/c SNe.

In such a high density environment, emission could occur in permitted
\ion{Fe}{ii} multiplets (e.g., Turrato et al. 2000; Deng et al. 2004), 
which might contribute to the blue continuum flux 
at $\lambda \lesssim 6000$ \AA .
The absence of intermediate-width [\ion{Fe}{ii}] line around 5200 \AA\ ,
which is seen together with [\ion{Fe}{ii}] 7155 in Type Ia SNe 
as well as in the peculiar Type Ic SN 1998bw 
(Fig. \ref{fig:SPECcomp3}), might be explained by this
higher density assumption.

The high density of the ejecta is also consistent with the 
current understanding
that the \Nifs\ heating model requires a massive progenitor 
and a large amount of ejecta (Umeda \& Nomoto 2008).
A large ejecta mass is also needed to reproduce the long 
duration of the LC around maximum.
The high density of the ejecta would also be consistent with
the lower expansion velocity ($\sim 1000$ \kms ) derived 
from FWHM of the emission lines in the late time spectrum.
Low expansion velocity was also suggested from early 
spectroscopy (Ofek et al. 2007).

\subsection{Unusual Emission Lines: From Inner Ejecta?}
\label{subsec:unusuallines}

There are several unusual features in the late time spectrum 
of SN 2006gy (marked with dashed lines in Fig. \ref{fig:SPECobs2}).
These lines are not clearly seen in other types of SNe 
(Fig. \ref{fig:SPECcomp1}--\ref{fig:SPECcomp3}).
Consulting line lists in literature (e.g., Moore 1945), 
we find possible candidates as shown in Table \ref{tbl:unident_line}.
Some of them could be lines of nickel
([\ion{Ni}{ii}] 7380, 8704; [\ion{Ni}{i}] 7394, 7908, 8202, 8843)
and/or titanium ([\ion{Ti}{ii}] 7917, 8040, 8060, 8349, 8371).
The emission line at 7716 \AA\ could be due to iron ([\ion{Fe}{i}] 7709 
or [\ion{Fe}{ii}] 7765).

If these identifications are correct, the innermost ejecta of 
SN 2006gy was visible at $t=394$ days because Ni and Ti are likely to 
exist only inner part of the ejecta.
Assuming that the line emitting region expands homologously with
$1000$ km s$^{-1}$ and all the elements are singly ionized, the high
density (
$10^{8}$ cm$^{-3} \lesssim n_{e} \lesssim 10^{9}$ cm$^{-3}$)
implies that the mass of the hypothetical innermost region
is $\sim$0.8--8$\Msun$ (if dominated by Fe-peak elements) or 
$\sim$0.3--3$\Msun$ (if dominated by Ca).
This would suggest that the assumption of $n_{e} \lesssim 10^{9}$ 
cm$^{-3}$ is reasonable.
In this case, the electron scattering optical depth is $\sim$0.2--2,
which suggests that $n_{e} \lesssim 10^{9}$ is also necessary for 
the innermost region to be optically thin.
In sum, our hypothesis of the emergence of the innermost part
requires $n_{e} \lesssim 10^{9}$ cm$^{-3}$, being consistent with 
the high density interpretation.

However, the identification of these lines are still tentative. 
For example, we do not see the [\ion{Ni}{i}] 7507 feature
in the observed spectrum (Fig. \ref{fig:SPECobs2}),
although it is expected to emit as strong as the [\ion{Ni}{i}] 7908.
It is noted the feature around 7910 \AA\ is also seen in some 
Type Ia SNe (Fig. \ref{fig:SPECcomp3}).
There is also an overall resemblance with the peculiar Type Ia 
SN 2005hk at 240 days, especially at 5500--6000 \AA\ ,
7100--7500 \AA\ and 8500--8900 \AA\ (Fig. \ref{fig:SPECcomp3}). 
To discuss this subject further, a more reliable identification 
of these lines will be needed.

\section{Conclusions}
\label{sec:conclusions}

We presented the spectroscopic and photometric observation 
of luminous Type IIn SN 2006gy at $t=394$ days.
The good seeing condition enabled us to detect it in 
the optical wavelengths.
The overall LC is roughly consistent with the radioactive heating 
model. The deviation between the observed LC and the radioactive 
LC model around 200 days suggests that the strong CSM interaction 
considerably contributed at $\sim$ 200 days and then weakened
by 394 days.

The late time spectrum of SN 2006gy was unique.
It showed some emission lines with intermediate width
($\sim 2000$ \kms ), including forbidden lines of [\ion{Fe}{ii}] 
and [\ion{Ca}{ii}] which are not seen in normal Type IIn/IIa SNe.
The absence of [\ion{O}{i}] 6300, 6364 line and the weakness of 
[\ion{Fe}{ii}] lines would be a result of moderate electron 
density of $\simeq 10^{8}$--$10^{9}$ cm$^{-3}$ in the 
emission line region.
This is consistent with the larger amount of ejecta and 
less expansion velocity suggested for this SN.
Although SN 2006gy exhibited possible H$\alpha$ emission line,
its flux was considerably lower than those of Type IIn/IIa
SNe and rather comparable with that of a typical Type II SN.
This suggests that the strong CSM interaction has finished
by $t=394$ days, being consistent with the prediction from the LC.

Several unusual emission features were present at 7400--8800\AA\ and
some of them might be Ni and/or Ti lines.
However, these identifications are still tentative and we need 
detailed modeling to explain the full aspects of the observation.

\acknowledgements

We are grateful to an anonymous referee for many helpful comments.
M.T. would like to thank Alex Filippenko, Weidong Li and 
Sergei Blinnikov for fruitful discussion and useful comments.
We have utilized SUSPECT database and would like to thank the 
managers of this database and all the contributors of the data 
used in the paper.

This research has been supported in part by the Grant-in-Aid for 
Scientific Research (17684004, 18104003, 18540231, 20740107, 20840007)
from the JSPS, MEXT, and World Premier International
Research Center Initiative, MEXT, Japan.
M.T. and N.T. are supported through the JSPS (Japan Society for the 
Promotion of Science) Research Fellowship for Young Scientists.

\begin{deluxetable}{ccccccccc}
  \tabletypesize{\scriptsize}
  \tablecaption{Log of the spectroscopic observations\label{tbl:obs}}
  \tablewidth{0pt}
  \tablehead{
   \colhead {} & \colhead{} & \colhead{} &
   \colhead{Res.} & \colhead{Exposure} & 
   \colhead{seeing} & \colhead{Airmass} & \colhead{} & \colhead{}\\
   \colhead{Epoch\tablenotemark{a}} & \colhead{UT} & \colhead{MJD} &
   \colhead{$\lambda / \Delta\lambda$} & \colhead{(s)} & 
   \colhead{(FWHM)} & \colhead{} & \colhead{$V$ mag} & \colhead{$R$ mag}
  }
  \startdata
   127 days & 2006 Dec 25.4 & 54094.4 & 3600 & 1200 &
    $1\farcs 0$--$1\farcs 1$ & 1.18--1.23 & $16.38\pm 0.1$ & $15.38\pm 0.1$\\
   157 days & 2007 Jan 24.4 & 54124.4 & 3600 & 1200 &
    $1\farcs 8$--$2\farcs 2$ & 1.68--1.84 & ---           & ---            \\
   394 days & 2007 Sep 18.5 & 54361.5 &  660 & 1200 &
    $0\farcs 5$--$0\farcs 6$ & 1.12--1.15 & $20.7\pm 0.4$ & $19.4\pm 0.4$\\
\enddata
\tablenotetext{a}{Rest frame days after explosion, 53967.5 MJD.}
\end{deluxetable}

\begin{deluxetable}{cccl}
  \tabletypesize{\scriptsize}
  \tablecaption{Emission lines on 394 days\label{tbl:unident_line}}
  \tablewidth{0pt}
  \tablehead{
   \colhead {Line center\tablenotemark{a}} & 
   \colhead{Intensity\tablenotemark{a}} &
   \colhead{FWHM\tablenotemark{a}} & \colhead{Comment}\\
   \colhead{(\AA )} & \colhead{($10^{15}$ erg s$^{-1}$ cm$^{-2}$} & 
   \colhead{(\AA )} & \colhead{} }
  \startdata
   6563?  & $\sim$1\tablenotemark{b} & 
    $\sim 20$\tablenotemark{b} & H$\alpha$ 6563\\
   7059 & $1.6\pm 0.6$   & $58\pm 8$ & \ion{He}{i} 7067\\
   7155 & $5.5\pm 2.2$   & $94\pm 6$ & [\ion{Fe}{ii}] 7155\\
   7302 & $6.0\pm 1.7$   & $83\pm 9$ & [\ion{Ca}{ii}] 7291, 7323,
   [\ion{O}{ii}] 7320, 7330, \ion{Fe}{ii} 7308\\
   7387 & $0.8\pm 0.8$   & $40\pm 16$ & [\ion{Ni}{ii}] 7380?, [\ion{Ni}{i}] 7394?\\
   7443 & $4.5\pm 0.5$   & $113\pm 15$ & \ion{Fe}{ii} 7462\\
   7716 & $2.8\pm 0.6$   & $117\pm 13$ & \ion{Fe}{ii} 7712, [\ion{Fe}{i}] 7709?, [\ion{Fe}{ii}] 7765?\\
   7913 & $2.3\pm 0.6$   & $57\pm 7$ & [\ion{Ni}{i}] 7908?, [\ion{Ti}{ii}] 7917?\\
   8052 & $6.0\pm 1.5$   & $93\pm 3$ & [\ion{Ti}{ii}] 8040?, [\ion{Ti}{ii}] 8060?\\
   8206 & $1.3\pm 0.5$   & $55\pm 6$ & [\ion{Ni}{i}] 8202?, [\ion{Ni}{i}] 8195?\\
   8325 & $1.9\pm 0.4$   & $62\pm 12$ & [\ion{Ti}{ii}] 8349?,[\ion{Fe}{i}] 8348?\\
   8383 & $1.9\pm 0.4$   & $49\pm 8$ & [\ion{Ti}{ii}] 8371\\
   8479 & $1.3\pm 0.3$   & $56\pm 10$ & \ion{Ca}{ii} 8498, [\ion{Fe}{i}] 8490?\\
   8544 & $4.8\pm 1.2$   & $56\pm 4$ & \ion{Ca}{ii} 8542\\
   8681 & $5.5\pm 0.9$   & $100\pm 7$ & \ion{Ca}{ii} 8662, [\ion{Ni}{ii}] 8704?\\
   8835 & $3.6\pm 0.4$   & $98\pm 7$ & [\ion{Ni}{i}] 8843?\\
\enddata
\tablenotetext{a}{Center wavelength, intensity and full-width at 
half-maximum of de-blended gaussian profiles from the SN spectrum 
are shown, except
for H$\alpha$. In the de-blending the continuum level was set to zero.
The errors are estimated from results of several de-blending trials for the
SN spectrum (in which the mean of scaled BG1 and BG2 is subtracted)
and also for other spectra (in which either only scaled BG1 or 
BG2 is subtracted; see Fig. \ref{fig:SPECobs2}).}
\tablenotetext{b}{Assumed a constant flux $0.61\times 10^{-16}$
erg s$^{-1}$ cm$^{-2}$ \AA $^{-1}$ (above the continuum level
of $0.52\times 10^{-16}$ erg s$^{-1}$ cm$^{-2}$ \AA $^{-1}$) 
and a total width of $\sim 20$ \AA (see \S \ref{sec:spec}).}
\end{deluxetable}

\end{document}